\documentclass{MYws-ijmpa}

\begin{document}

\markboth{T.E.O. Ericson, B. Loiseau, S. Wycech}
{A phenomenological $a^h_{\pi^-p}$ from pionic hydrogen}

\catchline{}{}{}{}{}
%
%%%%%%%%%%%%%%%%%%%%%%%%%%%%%%%%%%%%%%%%%%%%%%%%%%%%%%%%%%%%%%%%%%%%

\title{A PHENOMENOLOGICAL DETERMINATION OF THE PION-NUCLEON SCATTERING LENGTHS FROM PIONIC HYDROGEN\footnote{Invited talk at the $10^{\mathrm{th}}$ International Symposium on Meson-Nucleon Physics and the Structure of the Nucleon (MENU 2004), Beijing, China, August 29 - September 4, 2004 $-$ CERN-PH-TH/2004-227 $-$ LPNHE 2004-13}
}

\author{T.E.O. ERICSON}

\address{Theory Division, Physics Department,\\
 CERN, CH-1211 Geneva 23, Switzerland}

\author{\footnotesize B. LOISEAU}

\address{LPNHE\footnote{Unit\'e de Recherche des Universit\'es Paris 6 et Paris 7, associ\'e au CNRS}, Groupe Th\'eorie, Universit\'e P. \& M. Curie,\\
 4 Pl. Jussieu, F-75252 Paris, France
}

\author{S. WYCECH}

\address{Soltan Institute for Nuclear Studies,\\
PL - 00681 Warszawa, Poland
}

\maketitle

%\pub{Received (Day Month Year)}{Revised (Day Month Year)}

\begin{abstract}
A model independent expression for the electromagnetic corrections to a phenomenological hadronic pion-nucleon ($\pi N$) scattering length $a^h$, extracted from pionic hydrogen, is obtained. 
In a non-relativistic approach and using an extended charge distribution, these corrections are derived up to terms of order $\alpha^2\log\alpha$ in the limit of a short-range hadronic interaction. 
We infer $a^h_{\pi^-p}=0.0870(5)m^{-1}_\pi$ which gives for the $\pi NN$ coupling through the GMO relation $g^2_{\pi^\pm pn}/(4\pi)=14.04(17)$.

\end{abstract}

\vspace*{-0.3cm}
\section{Introduction}  
Precise knowledge of the strong interaction amplitudes at zero energy is important.
It gives strong constraints to the chiral physics approach\cite{gass02} to QCD at low energy.
Zero-energy amplitudes can also be used as subtraction constants in dispersion relations, as for instance in the zero energy pion-nucleon forward dispersion relation, the GMO sum rule, which allows a precise determination of the $\pi NN$ coupling\cite{eric02}.
Accurate knowledge of zero energy amplitudes requires careful analysis of electromagnetic corrections of precise experimental information from hadronic atoms.
For pionic hydrogen recent experimental results give for the strong interaction energy shift\cite{schr01,gott03}
 \begin{equation}
\label{eq:1}
\epsilon_{1s}= \left\{
\begin{array}{l}
       \big(-7.108 \pm 0.013 \mbox{ (stat)} \pm 0.034 \mbox{ (syst)}\big)\mbox{ eV } (\mbox{precision\cite{schr01}:}\pm 0.5\%)   \\
       \big(-7.120 \pm 0.008 \mbox{ (stat)} \pm  { 0.009\atop 0.008} \mbox{ (syst)}\big) \mbox{ eV } (\mbox{precision\cite{gott03}:}\pm 0.2\%) 
\end{array}
\right.
\end{equation}
and for the total decay width\cite{schr01}
\begin{equation}
\label{eq:G1s}
\Gamma_{1s}=[0.868\pm 0.040\mbox{ (stat) }\pm 0.038 \mbox{ (syst) } ]\mbox{ eV }(\mbox{precision\cite{schr01}:}\pm 6\%).
\end{equation}
It is well known\cite{dese54,true61}
\begin{equation}
\label{eq:E1s}
\epsilon_{1s}/E_B=(\epsilon^0_{1s}/E_B)(1+\delta_s)=4m\alpha\ a(1+\delta_{1s}).
\end{equation}
Here $E_B=-m\alpha^2/2$ is the Bohr energy, $\alpha=1/137.036$ the fine structure constant, $m$ the $\pi^-p$ reduced mass, $m=m_{\pi^-}m_p/(m_{\pi^-}+m_p)$.
The correction $\delta_{1s}$ is the deviation from the lowest order shift
\begin{equation}
\label{eq:eps0}
\epsilon^0_{1s}=-(4\pi/2m)\phi^2_B(0)a.
\end{equation}
The function $\phi_B(r)$ is the non-relativistic $1s$ Bohr wave function of a point charge.
The scattering length $a$ corresponds to the elastic threshold scattering amplitude defined in the absence of the Coulomb field.

It is important to understand $\delta_{1s}$ with an accuracy matching the high experimental precision.
The standard approach\cite{sigg96} uses a numerical resolution of coupled channel equations. 
The derivation is model dependent and not fully consistent with the $\pi N$ low-energy expansion\cite{eric02}.
It gives $\delta_{1s}(Sigg)=(-2.1\pm 0.5)\%$.
Calculation have also been performed in the framework of QCD+QED with effective field theory (EFT) techniques.
The correction from isospin symmetric pure QCD, evaluated within Chiral perturbation Theory (ChPT), is $\delta_\epsilon=(-4.3\pm 2.8)\%$ at leading order\cite{lyub00} and $\delta_\epsilon=(-7.2\pm 2.9)\%$ at next to leading order\cite{gass02}.
The large uncertainty in $\delta_\epsilon$ is due to the lack of knowledge on one of the low-energy constant entering in the EFT.
The above determinations consider $\delta_{1s}$ as corrections to an isospin symmetric world.

In the present work we study the relation between the strong atomic energy shift to the scattering length $a^h$ defined as the one which would be obtained if the Coulomb field of the extended charge could be withdrawn and considered as coming from an external source.
We give here the main features and results of our model, more details can be found in Ref.~\refcite{eric04}.
If the $1s$ finite size electromagnetic (e.m.) binding energy is denoted by $E_{fs}$ and the total $1s$ binding with strong interaction and finite size by $E$ then the strong interaction shift is $\epsilon_{1s}=E-E_{fs}$.
The non-relativistic wave number are $\kappa_B,\ \kappa_{fs}$ and $\kappa$ for $E_B,\ E_{fs}$ and $E$ respectively.
We have $\kappa^2_B=2mE_B$ and $\kappa^2=2mE$.
We also use the Bohr radius $r_B=\kappa_B^{-1}=(m\alpha)^{-1}$.

\vspace*{-0.3cm}
\section{Model for the $\pi^-p$ Atom}
In our approach the isospin symmetry is not assumed. 
The $\pi^-p$ atom being non-relativistic we consider our problem as a non-relativistic quantum one.
We work in the configuration space.
The pion-proton charge distributions are folded to give the Coulomb potential $V_c(r)$.
If $V_c(r)=0$, the low-energy expansion for the angular momentum $\ell=0$ is given in term of the phase shift $\delta_{\ell=0}$ and the momentum $q$ by
\begin{equation}
\label{eq:tan}
(\tan \delta_{\ell=0})/q=a^h+b^hq^2+...
\end{equation}
Here $a^h$ is the hadronic scattering length and $b^h$ the range parameter.

\vspace*{-0.2cm}
\subsection{Toy model}
We first consider a soluble toy model which will be extended to the realistic case.
We start by studying the case of a single channel.
We suppose the charge to be concentrated on a sphere of radius $R$ outside the range of the strong interaction.
Inside this shell, the Coulomb potential is constant with $V_C(R)=-\alpha/R\simeq 1.4$ MeV for a $R\sim 1$ fm. The $1s$ binding energy $E_B=-m\alpha^2/2\simeq -3.2$ keV is negligible compared to the Coulomb field and to the strong interaction inside the charge distribution region although its exact value governs the scale of the atom.
The inside wave function, $r\leq R$, is a standing wave outside the strong interaction region:
\begin{equation}
\label{eq:ur}
u(r)=N\left[\sin(q_cr)/q_c+\tan\delta^h_{\ell=0}\cos(q_cr)/q_c\right]
\end{equation}
where the wave number $q^2_c=2m\alpha/R-\kappa^2\simeq 2m\alpha/R$.
The external, $r\leq R$, $1s$ wave function is a Whittaker function,
\begin{equation}
\label{eq:ur2}
u(r)=(4\pi)^{-1/2}\kappa e^{-\kappa r}
\left\{
2\kappa r[(1+(1-\lambda)(1-\gamma-\log 2\kappa r)]+(1-\lambda)/\lambda
\right\}
\end{equation}
where $\lambda=\kappa_B/\kappa$ and $\gamma=0.577...$ is the Euler constant.
In Eq. (\ref{eq:ur2}) we have neglected terms of order 
$(1-\lambda)^2\simeq (\alpha ma^h)^2\simeq 10^{-6}$. 
Matching the logarithmic derivative of the wave function (\ref{eq:ur}) and (\ref{eq:ur2}) at $R$ gives, replacing $a$ by $a^h$ in Eq. ({\ref{eq:E1s}),
\begin{equation}
\label{eq:d1s}
\delta_{1s}=-2R/r_B+2(a^h/r_B)
[2-\gamma-\log(2\alpha mR)]+ (2m\alpha/R)(b^h/a^h)
\end{equation} 
In the absence of strong interaction, the extended charge wave function at $r=0$ is
$\phi_{in}(O)=\phi_B(O)(1-R/r_B+...)$ to the present order in $\alpha$.
It is a better e.m. starting function than $\phi_B(O)$ in Eq. (\ref{eq:eps0}).
This explains the first term in Eq. (\ref{eq:d1s}). 
The second term is a renormalization coming from the external wave function modified at $R$ by the hadronic scattering by a factor $1+2a^hm\alpha[2-\gamma-\log (2\alpha mR)]$.
This factor is insensitive to $R$.
The third term is a new effect, it arises from using the correct energy at the point of interaction.
It follows from gauge invariance\cite{eric02,eric04}.

\vspace*{-0.2cm}
\subsection{Generalization}

Any strong interaction, with the same near threshold hadronic amplitude and with a hadronic range inside $R$, will give the same answer.
The difference between the true charge distribution and that of the toy model is small and gives a perturbative potential.
We apply this perturbation to our soluble model to obtain\cite{eric04} an e.m. correction which is independent of $R$ to the present order in $\alpha$.
In  Eq. (\ref{eq:d1s}) $R$ is replaced by $\langle r\rangle_{em}$, $1/R$ by $\langle 1/r\rangle_{em}$ and $\log(mR)$ by $\langle \log(mR)\rangle_{em}$.
This result could also have been obtained by a direct calculation. 
The vacuum polarization correction $\delta^{vp}$ arises from a long-range potential which modifies the wave function at the origin.
This model independent correction has been calculated in Ref.~\refcite{eira00} to be $\delta^{vp}=0.48\%$.
It agrees with the earlier numerical value obtained in Ref.~\refcite{sigg96}.

\vspace*{-0.2cm}
\subsection{Coupled channel}

In addition to the $\pi^-p$ channel, the $\pi^-p$ atom couples to the $\pi^0n$ and $\gamma n$ channels.
We describe below the two-channel case with a K-matrix formalism.
Let us denote the charged $\pi^-p$ channel by $c$ and the neutral one by $o$.
The complex Coulomb threshold amplitude can be written as 
$A^c_{c,c}=K^c_{c,c}+iq_o(K^c_{c,o})^2$ where $q_0$ is the momentum in the $\pi^{0}n$ channel.
The bound state condition leads to\cite{true61},
\begin{equation}
\label{eq:e1s2}
\epsilon_{1s}-i\Gamma_{1s}/2=-(4\pi/2m)\phi^2_B(O)A^c_{c,c}
\left[1+2A^c_{c,c}(1+\gamma)/r_B\right]
\end{equation} 
The low-energy expansion for the hadronic K-matrix gives
$K_{c,c}=a^h_{c,c}+b^h_{c,c}q^2_c,\ K_{c,o}=a^h_{c,o}+b^h_{c,o}(q^2_c+q^2_o)/2$
and $K_{o,o}=a^h_{o,o}+b^h_{o,o}q^2_o$.
Continuity of the wave function matrix and its logarithmic derivative at $R$, together with the extended charge distribution, lead to
\begin{equation}
\label{eq:mat}
\begin{array}{rcccccc}
      \delta_{1s} & = & -2\displaystyle\frac{ \langle r\rangle_{em}}{ r_B} & + & 2 \displaystyle\frac{a^h}{r_B}\left (2-\gamma-\left\langle \log \displaystyle\frac{2}{r_B}\right\rangle_{em}\right) & + 
      & 2m\alpha \left \langle \displaystyle \frac{1}{r}\right \rangle_{em}\displaystyle \frac{b^h}{a^h}     \\
      \noalign{\medskip}
      &  \equiv  &\delta^{\langle r\rangle} & + & \delta^c & + & \delta^g 
\end{array}
\end{equation}
\begin{equation}
\label{eq:Gpi}
\rm{and}\ 
\Gamma^{\pi^on}_{1s}=\Gamma_{1s}/(1+P^{-1})=(4\pi/m)\phi^2_B(o)q_o
\left[a^h_{c,o}(1+\delta_\Gamma)\right]^2
\end{equation}
with $\delta_\Gamma\equiv \delta^{\langle r\rangle}/2+\delta^c+(q^2_c+q^2_o)b^h_{c,o}/(2a^h_{c,o})+\delta^{vp}/2$.
The Panofsky ratio\cite{spul77},  $P=1.546(9)$.
It is the ratio of the $\pi^0$ to $\gamma$ production  cross sections.

\vspace*{-0.3cm}
\section{Numerical Results}

The e.m. expectations values in Eq. (\ref{eq:mat}) are calculated using the folded $(\pi^-,p)$ charge distribution obtained from the observed form factors as in Ref.~\refcite{sigg96}.
The range terms values $b_{\pi^-p}=b_{\pi^+n}=-0.031(9)m^{-3}_\pi$ and
$b_{\pi^-n}=b_{\pi^+p}=-0.058(9)m^{-3}_\pi$ are from $\pi N$ scattering data\cite{hohl83}.
From Eqs. (\ref{eq:E1s}) and (\ref{eq:mat}), including $\delta^{vp}$ one has,
\begin{equation}
\label{eq:res}
\epsilon_{1s}=-(4\pi/2m)\phi^2_B(0) a^h_{c,c}(1+\delta^{\langle r\rangle}+\delta^c+\delta^g+\delta^{vp}).
\end{equation}
Using then the experimental value $\epsilon_{1s}$ (\ref{eq:1}) of Ref.~\refcite{schr01}, plus a two-step iteration from a starting value of $a^h_{c,c}$, lead to the $\delta_{1s}$ results given in Table 1.
The main source of uncertainty comes from the empirical range parameters.
\begin{table}[h]
\tbl{Coulomb correction in percent as described in the text. $\delta^{vp}$ is added to $\delta_{1s}$.}
{\begin{tabular}{cccccc}\toprule
 &   Extended charge   &  Renormalization    &  Gauge term    & Vacu. Polar. & Total  \\
 &  $\delta^{\langle r\rangle}$ & $\delta^c$ & $\delta^g$ & $\delta^{vp}$ & \\
\colrule
 $\delta_{1s}$ & -0.853(8) & 0.701(4)    &  -0.95(29)  &  0.48  &  -0.62(29)             \\
 $\delta_\Gamma $ & -0.427(4) & 0.701(4) & 0.50(23) & 0.24 & 1.02(23)            \\
 \botrule
\end{tabular}}
\end{table}
The resolution of Eq. (\ref{eq:res}) and that of Eq. (\ref{eq:Gpi}), with the experimental value of $\Gamma_{1s}$ from (\ref{eq:G1s}), together with the corrections as given in Table 1, plus a negative sign for $a^h_{c,o}$, lead to
\begin{equation}
\label{eq:ah}
 a^h_{\pi^-p}     =  a^h_{c,c}  =  0.0870(5)m^{-1}_\pi \quad ; \quad
   a^h_{\pi^-p\to\pi^{o}n}     =   a^h_{c,o}  =  -0.0125(4) m^{-1}_\pi .
\end{equation}
The result for $a^h_{\pi^-p}$ is 1.5\% smaller and outside the quoted uncertainty of the calculated value $0.0883(8)m^{-1}_\pi$ of Ref.~\refcite{schr01} using Sigg analysis\cite{sigg96}.
Our value of $a^h_{\pi^-p\to\pi^{o}n}$ is smaller by 2.4\% from that of Ref.~\refcite{schr01}, 
$-0.128(6)m^{-1}_\pi$.
The $a^h_{\pi^-p}$ can be analyzed together with $a_{\pi^-D}$ to give an isovector scattering length $(a_{\pi^-p}-a_{\pi^-n})/2$.
We follow Ref.~\refcite{eric02} with two small additions: we consider the full triple scattering correction\cite{dori04,bean03} and the additional Fermi motion correction arising from the energy dependence of the S-wave isoscalar amplitude at threshold, as first discovered in a chiral approach\cite{bean03}.
One obtains $(a_{\pi^-p}-a_{\pi^-n})/\sqrt{2}=-0.125(1)m^{-1}_\pi$ in perfect agreement with the value given in (\ref{eq:ah}).
Note here that, in the limit of charge symmetry, the further e.m. contributions, for instance from processes such as  the $\gamma N(\gamma\Delta)$ channels and their cross terms, cancel.
There is no evidence of isospin violation in the isovector amplitude at the current level of accuracy.

Our result for $a^h_{\pi^-p}$, together with the reanalysis of the $\pi^-D$ scattering length, as just described above, allows an improved determination of the $\pi NN$ coupling constant via the GMO sum-rule as given in Eq. (4) of Ref.~\refcite{eric02}.
We obtain $g^2_{\pi^\pm pn}/(4\pi)=14.07(17)$ to be compared to 14.11(19) in Ref.~\refcite{eric02}.

\vspace*{-0.3cm}
\section{Comparison to Previous Approaches and Conclusions}

In former analytical approaches\cite{dese54,true61,carb92,hols99} using wave functions, the effect of the e.m. finite size and the issue of the correct energy of the interaction were basically not taken into account.
The correction proportional to $(a^c_{c,c})^2$ is obtained to leading order $\alpha\log\alpha$, but the part of order $\alpha$ is incorrectly given, leading to a small numerical difference.
The numerical approach of Ref.~\refcite{sigg96}, with coupled channel Klein-Gordon equations, includes correctly the e.m. finite size, the vacuum polarization and the renormalization term.
It does not have the proper range parameters and makes model dependent corrections for isospin violation and radiative effects.
The numerical result for a single channel does not have these problems and agrees with our result when using the same input.

The energy shift $\epsilon_{1s}$ is related to the scattering length in pure QCD in Refs.~\refcite{gass02} and \refcite{lyub00}.
The calculations are done in a perturbative expansion to the same order in $\alpha$ as here in an effective quantum field theory framework.
The e.m. EFT corrections for mass splitting are implicitly contained in our $a^h$.
One cannot then compare in detail both expansions.
Our renormalization term proportional to $(a^h)^2$ is also present in the EFT approach.
The main difference is the wave function and energy shift corrections linear and inverse in the e.m. charge radius, respectively.
No such terms seem to be present in the EFT in which only even power $\langle r^{2n}\rangle_{em}$ appear.
It is due to a different treatment of the e.m. charge form factor\cite{eric04}.
We conjecture that the EFT approach should produce terms corresponding to ours as form factors are generated using such descriptions.

Let us conclude by the following important observation.
The Coulomb potential for the extended charge distribution is regular at short distance and it plays the role of an externally applied binding potential in addition to the strong interaction.
The problem can then be solved exactly for a model where perturbation can be applied.
We achieve an accurate determination of the correction terms to the Deser-Trueman formula 
(\ref{eq:E1s}) in a non-relativistic picture with three quite understood physical effects.
First the wave function at the origin should match with an extended charge distribution including vacuum polarization.
Second, the correct long-range comportment of the wave function changes in a characteristic way the wave function at the origin.
It leads to a correction proportional to $\alpha m(a^h)^2\log\alpha$ to leading order.
Thirdly the low-energy expansion of the scattering amplitude gives a characteristic "gauge" correction\cite{eric02}.
This effect is as important as the others.
Our results should not change as long as the hadronic range is within that set by the charge distribution.
Our approach is general and can be easily applied to other atomic systems like $\pi^-\pi^+$.

\end{document}